\documentclass[pra,twocolumn,showpacs,preprintnumbers,amsmath,amssymb]{revtex4-1}
\usepackage{graphicx}% Include figure files
\usepackage{dcolumn}% Align table columns on decimal point
\usepackage{bm}% bold math
\usepackage{braket}% Dirac notation
\usepackage{comment}% block comment option

\usepackage{color} 				%% colored text
\usepackage[normalem]{ulem} 			%% crossed text
\begin{document}

\title{Critical entropies and magnetic-phase-diagram analysis of ultracold three-component fermionic mixtures in optical lattices
}

\author{Andrii Sotnikov}
\affiliation{Akhiezer Institute for Theoretical Physics, NSC KIPT, 61108 Kharkiv, Ukraine}
\affiliation{Karazin Kharkiv National University, 61022 Kharkiv, Ukraine}
%\affiliation{Institut f\"ur Theoretische Physik, Goethe-Universit\"at, 60438 Frankfurt/Main, Germany}

\date{\today}% It is always \today, today,

%% ABSTRACT %%
\begin{abstract}
 We study theoretically many-body equilibrium magnetic phases and corresponding thermodynamic characteristics of ultracold three-component fermionic mixtures in optical lattices described by the SU(3)-symmetric single-band Hubbard model. Our analysis is based on the generalization of the exact diagonalization solver for multicomponent mixtures that is used in the framework of the dynamical mean-field theory. It allows us to obtain a finite-temperature phase diagram with the corresponding transition lines to magnetically ordered phases at filling one particle per site (1/3 band filling) in simple cubic lattice geometry. Based on the developed theoretical approach, we also attain the necessary accuracy to study the entropy dependence in the vicinity of magnetically ordered phases that allows us to make important predictions for ongoing and future experiments aiming to approach and study long-range-order phases in ultracold atomic mixtures.
\end{abstract}

%, 71.45.Lr -- CDW systems, 75.50.Ee -- antiferromagnetism; 75.10.Jm --
\maketitle

%% INTRODUCTION %%
\section{Introduction}
There has been remarkable progress in approaching quantum magnetism in ultracold two-component atomic mixtures in optical lattices, where short-range antiferromagnetic correlations were detected and measured recently \cite{Greif2013S,Har2015Nat}. It is believed that observations and subsequent detailed studies of the related long-range-order states of matter, together with other novel quantum phases, will help not only in designing new generations of strongly correlated materials, but also in quantum simulation applications.
At the same time, the mentioned experimental success, together with still existing challenges (e.g., limitations in cooling \cite{Jor2010PRL}), shifts the research horizons further. 

From a theoretical point of view, while the main ingredients for long-range-order magnetic phases (e.g., phase diagrams and critical entropies) are determined with a high accuracy for the single-band two-component fermionic Hubbard model (see Refs.~\cite{Kent2005PRB,Fuchs2011PRL} for details), there are many fewer quantitative predictions for multicomponent mixtures.
However, there is clear evidence that these systems possess a very rich physics that results in a variety of quantum many-body phases and their unique properties \cite{Che2007PRL,Rap2008PRB,Tot2010PRL,Rap2011PRA,Szi2011EPL,Ina2013MPL,Sotnikov2014PRA}.

In this paper we present a theoretical approach that allows us to study long-range-order magnetic phases in the SU(3)-symmetric Hubbard model in more detail and determine the finite-temperature phase diagram structure depending on the main system parameters, the hopping amplitude and the interaction strength, that can be tuned with a high degree of freedom for ultracold atoms in optical lattices. However, we not only aim to bridge the mentioned gap, but also analyze by this example whether multicomponent mixtures can be considered advantageous in comparison with their two-component counterparts for the purpose of approaching specific long-range-order states as suggested in Refs.~\cite{Tai2012Nat,Haz2012PRA,Cai2013PRL}.

%% SYSTEM AND MODEL %%
\section{System and Model}\label{sec.2}
We consider a system described by the SU(3)-symmetric Hubbard Hamiltonian of the type
\begin{eqnarray}\label{HubbSU3}
\mathcal{\hat{H}}=&&
-t\sum\limits_{\langle ij\rangle}\sum\limits_{\alpha=1}^3 \bigl( \hat{c}^\dag_{i\alpha}\hat{c}_{j\alpha}+{\rm H.c.}\bigr)
+U\sum\limits_{i}\sum\limits_{\beta>\alpha}^3\sum\limits_{\alpha=1}^3\hat{n}_{i\alpha}\hat{n}_{i\beta}
\nonumber\\
&&+\sum\limits_{i}(V_i-\mu)\hat{n}_{i},
\end{eqnarray}
where $t$ is the hopping amplitude between nearest-neighbor sites $i$ and $j$ with the corresponding shorthand notation $\langle ij\rangle$ under the sum, $U$ is the on-site interaction strength between atoms in different internal (e.g., hyperfine) states $\alpha$ and $\beta$, $\hat{c}^\dag_{i\alpha}$ ($\hat{c}_{i\alpha}$) is the fermionic creation (annihilation) operator of the atom in the state~$\alpha$ on the lattice site~$i$, and $\hat{n}_{i\alpha}=\hat{c}^\dag_{i\alpha}\hat{c}_{i\alpha}$ and $\hat{n}_{i}\equiv\sum_{\alpha}\hat{n}_{i\alpha}$ denote operators of the particle number. In the second line, we also introduced the amplitude of the external trapping potential $V_i$ on the lattice site $i$ and the chemical potential $\mu$ that sets the total number of atoms in the system. We assume that the optical lattice is deep enough $V_{\text{lat}}\gtrsim 5E_r$ ($E_r$ is the recoil energy of atoms) for the single-band approximation to be correct.

The system described by the Hamiltonian (\ref{HubbSU3}) can be experimentally realized by loading ultracold atoms that are prepared in three different quantum hyperfine states to the optical lattice. Since atoms in all three internal states (called spin components below) must interact with each other with the same strength $U$, appropriate candidates are mixtures of alkaline-earth atoms ($^{173}$Yb and $^{87}$Sr; see, e.g., Refs.~\cite{Tai2010PRL,Ste2011PRA}) and mixtures of $^{40}$K atoms, where by means of a proper choice of hyperfine states and Feshbach resonances one can realize a regime with almost equal scattering lengths between spin components. In both experimental cases the lattice depth can be easily tuned by the intensity of laser beams, thus the coupling $U/t$ can be changed in a wide range.

It is worth emphasizing that in the case of a homogeneous system ($V_i=0$) and filling one particle per site (i.e., $n=1$ or, equivalently, 1/3 band filling that is, strictly speaking, only approximately fulfilled by setting $\mu\approx U/2$; see Ref.~\cite{Sotnikov2014PRA} for details) in the strong-coupling limit $U/t\gg1$ and by using an established technique \cite{Mac1988PRB} one arrives at the SU(3)-symmetric Heisenberg model
\begin{eqnarray}\label{HeisSU3}
  \mathcal{\hat{H}}_{\textrm{eff}} = J
  \sum\limits_{\langle ij\rangle}
  \sum\limits_{k=1}^{8} \hat{S}_{ki}\hat{S}_{kj}
\end{eqnarray}
with positive (antiferromagnetic) exchange coupling $J=4t^2/U$ and the local pseudospin projection operators $\hat{S}_{ki} = \frac{1}{2} \hat{c}^\dag_{\alpha i} \lambda_{k\alpha\beta} \hat{c}_{\beta i}$, which are expressed in terms of $3\times3$ unitary Gell-Mann matrices $\boldsymbol{\lambda}_k=\{\boldsymbol{\lambda}_1,\ldots,\boldsymbol{\lambda}_8\}$ \cite{Georgi1999}.
Below, we do not limit ourselves to the strong-coupling limit; we study thermodynamic properties of the Hubbard model (\ref{HubbSU3}) at filling $n=1$ depending on the coupling $U/t$ and the temperature~$T/t$. At the same time, the effective Hamiltonian~(\ref{HeisSU3}) is useful for a proper analysis of particular effects obtained by the theoretical approach introduced below.

%% METHOD %%
\section{Method}\label{sec.3}
We use a dynamical mean-field theory (DMFT), the general idea and specific details of which can be found in Ref.~\cite{Geo1996RMP}. Regarding other state-of-the-art numerical methods, it is important to note that DMFT can be considered at present as one of the most optimal approaches for the system under study. In particular, the determinant quantum Monte Carlo method \cite{Suzuki1986} cannot be applied for the SU(3)-symmetric mixture due to the presence of the sign problem that cannot be suppressed or excluded as, for example, in the SU(4)-symmetric mixture at half filling \cite{Cai2013PRL}. At the same time, the direct exact diagonalization approach is very limited in the number of lattice sites for such systems \cite{Tot2010PRL} that may cause significant difficulties in a reliable finite-size scaling analysis for magnetic phase diagrams.
Therefore, below, we briefly introduce key generalizations and main steps in DMFT that allow us to obtain the central results of the paper and can be important for further studies of multicomponent mixtures in the framework of the approach used.

\subsection{Generalization of the exact diagonalization impurity solver for multicomponent mixtures}
The central idea of DMFT consists in mapping the original lattice problem (\ref{HubbSU3}) that is usually intractable to the local (impurity) problem that can be solved exactly. Taking the Anderson impurity model (AIM), one can notice that it can be extended to mixtures with a multiple number~${\cal N}$ of spin components in a straightforward way.
Since we use the exact diagonalization (ED) solver in our simulations (originally introduced in Ref.~\cite{Caffarel1994PRL} for two spin components ${\cal N}=2$), the corresponding impurity Hamiltonian can be written as
\begin{eqnarray}\label{HamAIM}
\hat{\cal H}_{\text{AIM}}
= & &
  \sum_{\sigma=1}^{\cal N}\sum_{l=2}^{n_s} \varepsilon_{l\sigma} \hat{a}_{l\sigma}^\dag \hat{a}_{l\sigma} 
+ \sum_{\sigma=1}^{\cal N}\sum_{l=2}^{n_s} V_{l\sigma} (\hat{a}_{l\sigma}^\dag \hat{d}_{\sigma} + {\rm h.c.})
\nonumber
\\
&+& U\sum_{\sigma'>\sigma}^{\cal N}\sum_{\sigma=1}^{\cal N}\hat{n}_{d\sigma}\hat{n}_{d\sigma'}
 - \mu\sum_{\sigma=1}^{\cal N}\hat{n}_{d\sigma},
\end{eqnarray}
where $\sigma,\sigma'=\{1,\ldots,{\cal N}\}$ denote the spin indices and the index $l=\{2,\ldots,n_s\}$ labels the number of the bath orbitals in the AIM with $n_s$ being the cutoff number peculiar to the ED approach. The two terms in the first line of Eq.~(\ref{HamAIM}) correspond to the energies of electrons in the bath and the hybridization between the bath and the impurity, respectively.
The operators $\hat{a}^\dag_{l\sigma}$ ($\hat{a}_{l\sigma}$) and $\hat{d}^\dag_{\sigma}$ ($\hat{d}_{\sigma}$) are the creation (annihilation) operators of electrons on the bath's orbital $l$ and the impurity, respectively; the quantities $\varepsilon_{l\sigma}$ and $V_{l\sigma}$ are the so-called Anderson parameters that set the amplitude of the processes in this model. These parameters are determined iteratively in the DMFT approach.
The terms in the second line have a direct correspondence to the original Hubbard Hamiltonian~(\ref{HubbSU3}) as a generalization to the SU($\cal N$)-symmetric mixture. 

In the Fock-space representation, the occupation of the orbital~$l$ by the spin component~$\sigma$ can take two values $n_{l\sigma}=\{0,1\}$, thus, in the most general case, the matrix that is necessary to diagonalize within the ED solver is of the size $L_0\times L_0$ with $L_0=2^{n_s{\cal N}}$. This results in a strong limitation of the number of orbitals $n_s$ that is feasible to account in numerical calculations for the given number of spin components ${\cal N}$. However, analogously to Refs.~\cite{Caffarel1994PRL, Geo1996RMP}, due to the structure of the Hamiltonian~(\ref{HamAIM}) that corresponds to the original lattice problem~(\ref{HubbSU3}), we note that in the case under study it does not mix different spin sectors, $|n_{11},\ldots, n_{n_s1}\rangle\cdots|n_{1{\cal N}}\ldots, n_{n_s{\cal N}}\rangle$, thus the total charge per spin component $q_\sigma=\sum_{l=1}^{n_s}n_{l\sigma}$ can be considered as a conserved quantum number. Defining the quantum state as $|Q_i\rangle\equiv|q_1^{(i)},\ldots,q_{\cal N}^{(i)}\rangle$, we determine in that way the total number of distinct states, $i=\{1,\ldots,(n_s+1)^{\cal N}\}$.

Therefore, the matrix $L_0\times L_0$ diagonalization problem transforms to diagonalizing $(n_s+1)^{\cal N}$ blocks of a different (but much smaller) linear size~$L_i$ expressed in terms of the binomial coefficients $L_i=\prod_{\sigma=1}^{\cal N}\binom{n_s}{q_\sigma^{(i)}}$. The described procedure allows us to account for more orbitals in the impurity problem, thus increasing the corresponding accuracy of results. In particular, from the direct calculation analysis, we conclude that by comparing to two-component mixtures with $n_s=7$, it is completely feasible to use $n_s=5$ for three-component and $n_s=4$ for four-component mixtures within the ED solver in DMFT approach. %Also note that the diagonalization procedure can be further improved, if necessary, by accounting the absence of hybridization within the bath orbitals in Eq.~(\ref{HamAIM}), thus introducing an additional subset of quantum numbers.

After the diagonalization is performed, we obtain the interacting Green's functions in the Matsubara-frequency space according to the standard definition (see also \cite{Geo1996RMP})
\begin{eqnarray}\label{eq.GF}
 G_{\sigma}(i\omega_n) = \frac{1}{\cal Z}\sum_{j,k}
 \frac{\langle j|\hat{d}_{\sigma}|k\rangle\langle k|\hat{d}^\dag_{\sigma}|j\rangle}{E_j-E_k-i\omega_n},
\end{eqnarray}
where ${\cal Z}=\sum_j e^{-E_j/T}$ is the partition function, the sum is taken over the full set of eigenstates $|j\rangle$ and $|k\rangle$ with the corresponding energies $E_j$ and $E_k$, $\omega_n = \pi(2n+1)T$ is the fermionic Matsubara frequency, and $T$ is the temperature (we use units such that $k_B=1$).
Subsequently, the local self-energies for each spin component are calculated from the Dyson equation
\begin{eqnarray}\label{Dyson}
 \Sigma_{\sigma}(i\omega_n) = {\cal G}^{-1}_\sigma(i\omega_n) - {G}^{-1}_\sigma(i\omega_n),
\end{eqnarray}
where the Weiss Green's function is defined by the set of Anderson parameters entering the Hamiltonian~(\ref{HamAIM}) through the analytic relation ${\cal G}^{-1}_\sigma(i\omega_n)=i\omega_n+\mu-\sum_{l=2}^{n_s} V_{l\sigma}^2/(i\omega_n-\varepsilon_{l\sigma}).$

\subsection{Clustering in the real-space lattice projection for homogeneous systems}

To obtain the lattice Green's functions for self-consistency conditions in DMFT, one can use the noninteracting density of states~$D(\epsilon)$ for a particular lattice geometry and the local self-energies~(\ref{Dyson}) obtained within the impurity solver. However, this method can be successfully applied (see, e.g., Ref.~\cite{Geo1996RMP} for more details when there is no translational invariance broken (e.g., in the paramagnetic or ferromagnetic phase),
\begin{eqnarray} \label{eq.1sub}
 G_{\sigma} (i \omega_{n}) = \int \frac{D(\epsilon)d\epsilon }{i \omega_{n} + \mu - \epsilon - \Sigma_{\sigma}(i \omega_n)}
\end{eqnarray}
or the symmetry is broken towards a bipartite structure (e.g. charge-density wave or antiferromagnetic N\'{e}el-type ordering)
\begin{eqnarray} \label{eq.2sub}
 G_{\sigma s}(i\omega_n) = 
 \int \frac{\zeta_{\sigma\bar{s}} D(\epsilon) d\epsilon}{\zeta_{\sigma A} \zeta_{\sigma B}-\epsilon^2}
\end{eqnarray}
where $\zeta_{\sigma s}\equiv i\omega_n + \mu - \Sigma_{\sigma s}(i \omega_{n})$, the sublattice indices $s = A, B$, and its opposite $\bar{s} = B, A$.

Obviously, there is no general expression possible in terms of $D(\epsilon)$ for more exotic types of magnetic order and, in particular, for the three-sublattice antiferromagnetic state that is peculiar to three-component mixtures at low temperatures \cite{Tot2010PRL,Sotnikov2014PRA}. Therefore, a real-space generalization of DMFT (RDMFT) (introduced in Refs.~\cite{Hel2008PRL,Sno2008NJP}) can be considered as a proper extension to account for more types of magnetic order in an unbiased way \cite{Sotnikov2014PRA}. Within this approach, the corresponding lattice Green's function is obtained from the inversion of the following real-space matrix:
\begin{eqnarray} \label{eq.Nsub}
 [{\bf G}^{-1}_{\sigma}(i\omega_n)]_{ ss'} = [i \omega_{n} + \mu - \Sigma_{\sigma s}(i \omega_n)]\delta_{ ss'} - t_{ ss'},
\end{eqnarray}
where the indices $s,s'$ denote the lattice sites, $\delta_{ ss'}$ is the Kronecker symbol, and $t_{ ss'}$ is the hopping matrix element that equals $t$ if there is tunneling possible between the lattice sites $s$ and $s'$ and zero otherwise, which depends on the lattice geometry, system size, and the boundary conditions (open or periodic) used in the original model~(\ref{HubbSU3}).

\begin{figure}
\includegraphics[width=\linewidth]{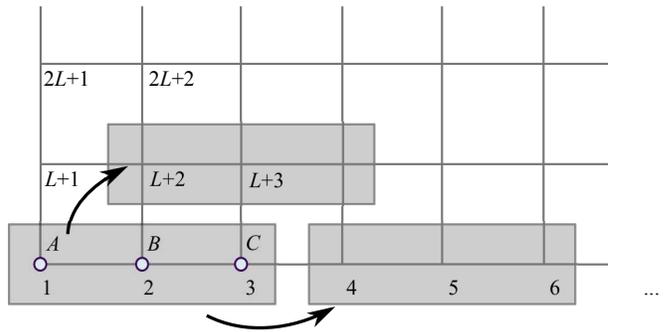}
    \caption{\label{fig1} (Color online)
    Schematic illustration of the clustering procedure in RDMFT by an example of the three-sublattice structure in a square lattice geometry ($L$ is the number of sites in the $x$ direction) that allows for the antiferromagnetic phase with the ordering wave vector ${\bf Q}=(2\pi/3,2\pi/3)$.}
\end{figure}

It is important to note that RDMFT requires more computational resources in comparison to the single-site (\ref{eq.1sub}) or two-sublattice (\ref{eq.2sub}) DMFT. It is caused not only by the necessity of the matrix inversion (which is feasible for systems consisting of $\sim\!\!10^4$ sites or even more), but also because the self-energies~$\Sigma_{\sigma s}(i \omega_n)$ must be calculated within the impurity solver on every lattice site~$s$ of the original lattice problem (\ref{HubbSU3}). However, the latter limitation can be significantly reduced for homogeneous systems in the case when the type of breaking of the translational symmetry is known beforehand (e.g., from unbiased RDMFT calculations for smaller system sizes). In Fig.~\ref{fig1}
we show one particular type of clustering that allows for the possibility of three-sublattice antiferromagnetic ordering in the system. This simplification requires the impurity problem to be solved only on three lattice sites, while the Green's functions are obtained from Eq.~(\ref{eq.Nsub}) for the full-size system, thus effectively minimizing the influence of the finite-size effects.

\subsection{Entropy analysis}
Concerning experiments with ultracold atomic mixtures in optical lattices, a crucial quantity for approaching magnetically ordered phases is not the temperature (which is a natural variable in the DMFT approach operating in the framework of the grand-canonical description) but the entropy, since atoms in optical lattices can be considered to some extent as isolated from the surrounding environment and key system parameters (e.g., the coupling strength $U/t$) can be changed adiabatically.
Therefore, quantitative theoretical predictions for the entropy become of a high importance in this field of research.

To calculate the entropy in our theoretical approach, according to Ref.~\cite{Wer2005PRL}, it is efficient to use the Maxwell relation for the entropy per lattice site $\partial s/\partial\mu =\partial n/\partial T$. It can also be written in the form
\begin{eqnarray} \label{eq.entr1}
 s(\mu,U,T)=\int_{-\infty}^{\mu}\frac{\partial n(\mu',U,T)}{\partial T}d\mu',
\end{eqnarray}
which is more suitable for the DMFT analysis, since $\mu$, $U$, and $T$ are input parameters, whereas the filling $n$ is the local observable measured within the impurity solver.

However, in the direct numerical analysis it is rather convenient to parametrize the chemical potential to the form $\mu(r) = \mu_1 - Vr^2$ that is analogous to the local-density approximation for our system in the external trapping potential of the parabolic shape with the amplitude $V$ and the radial distance $r$ measured in units of the lattice constant. Therefore, now we also access the entropy distribution in the harmonic trap, since the entropy at the given radial point $r_1$ is defined as follows:
\begin{eqnarray} \label{eq.entr2}
 s(r_1,U,T)=2V\int_{r_1}^{R_{\max}}\frac{\partial n}{\partial T}rdr,
\end{eqnarray}
with the boundary condition at the edge point~$R_{\max}$ of the trap $s(R_{\max})=0$ that is fulfilled in our calculations by the condition $n(r\geq R_{\max})=0$.

It is worth noticing that for the entropy analysis of homogeneous systems that is applied to the phase diagram with the fixed filling $n_1$ (e.g., $n_1=1$), the parameter $r_1$ in Eq.~(\ref{eq.entr2}) must be determined independently at every $(U,T)$ point from the condition $n(r_1,U,T)=n_1$ (see also Ref.~\cite{Sotnikov2014PRA}). Probably, the only exception can be made in this respect for the case of half filling in the SU($\cal N$)-symmetric Hubbard model, where one obtains the fixed condition for the chemical potential $\mu_{\text{hf}}=({\cal N}-1)U/2$ that is independent of quantum and thermal fluctuations due to particle-hole symmetry at any $\cal N$.

\section{Results}\label{sec.4}

\begin{figure}
\includegraphics[width=\linewidth]{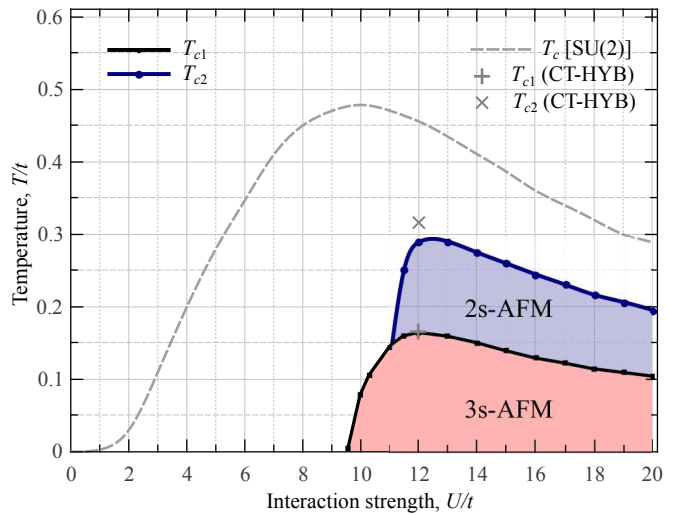}
    \caption{\label{fig2} (Color online)
    Finite-temperature phase diagram for magnetic phases in the SU(3)-symmetric Hubbard model at filling one particle per site ($n=1$) obtained by DMFT.
    At weak coupling and low temperatures ($T<0.1t$) the system size was chosen large enough ($\geq9^3$ lattice sites) to minimize the influence of finite-size effects and to guarantee reliable numerical convergence of RDMFT.
    The reference data for the points $T_{c1,2}$~(CT-HYB) is taken from Ref.~\cite{Sotnikov2014PRA}.}
\end{figure}

With the theoretical approach developed we study the magnetic ($T-U$)-phase diagram. Concerning the specific choice of the ED solver for this purpose in DMFT, we find that within the necessary accuracy it allows us to access low enough temperatures and has flexibility in the choice of the coupling strength $U/t$ in the Hubbard Hamiltonian~(\ref{HubbSU3}) even for $n_s=4$. Here, according also to the limitations of the chosen impurity model~(\ref{HamAIM}) and analogously to Refs.~\cite{Ina2013MPL,Sotnikov2014PRA}, we restrict ourselves to the case when the SU(3) symmetry can be spontaneously broken in two directions (i.e., along two easy axes that correspond to the natural-color basis) in the pseudospin space. In particular, we measure relative occupations by each spin component or, equivalently, analyze two local magnetizations on the lattice site~$s$,
\begin{eqnarray} \label{eq.magn}
 &&m^{(3)}_s\equiv \langle \hat{S}_{3s}\rangle = (n_{1s}-n_{2s})/2,
 \nonumber
 \\
 &&m^{(8)}_s\equiv \langle \hat{S}_{8s}\rangle = (n_{1s}+n_{2s}-2n_{3s})/\sqrt{3},
\end{eqnarray}
and their periodic behavior in real space. Therefore, magnetically-ordered states are determined from the convergence analysis for the Green's functions with different self-consistency conditions (\ref{eq.1sub})-(\ref{eq.Nsub}), which allow for different types of the sublattice structure (i.e., different antiferromagnetic phases). Hence, the main results are summarized in Fig.~\ref{fig2}.

By using the same scale in Fig.~\ref{fig2} we also show the transition line obtained by DMFT for the two-component SU(2)-symmetric mixture. This comparison directly shows a complete suppression of the antiferromagnetic order at weak coupling that we attribute to the absence of the key mechanism (leading to the so-called Slater-type antiferromagnet; see, e.g., Ref.~\cite{Fradkin2013}) caused by the absence of the perfect nesting of the Brillouin zone, since the band is only 1/3 filled. In the opposite case of strong coupling $U/t\gg1$, as we mentioned, there is a direct mapping to the effective spin model (\ref{HeisSU3}), thus the Heisenberg-type antiferromagnetic states arise at finite temperature in a cubic lattice geometry. This mechanism is also confirmed in the phase diagram by a corresponding decrease of the critical temperatures $T_{c1,2}$, which are proportional in both cases to the magnetic coupling $J=t^2/U$ in the limit $U/t\gg1$.

Concerning the specific spatial structure of the ordered states, we show their possible in-plane configurations in Fig.~3 (for the sake of simplicity we omit the third spatial dimension in illustrations since the corresponding extension is rather straightforward; see also Ref.~\cite{Sotnikov2014PRA}).
\begin{figure}
\includegraphics[width=\linewidth]{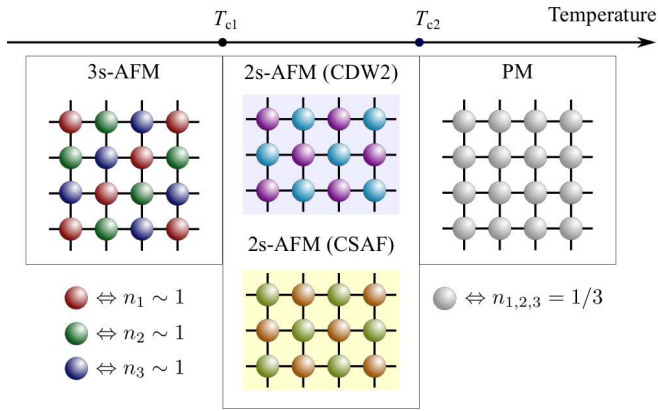}
    \caption{\label{fig3} (Color online)
    Schematic illustration of the observed magnetic phases at a filling of one particle per site. The color background indicates the presence of the additional net magnetization in the 2s-AFM states.}
\end{figure}
Note that other equivalent spatial arrangements for possible three-sublattice antiferromagnetic (3s-AFM) states can be obtained by $\pi/2$ rotations and spatial translations by the lattice constant along main axes. 
As for the two-sublattice antiferromagnetic (2s-AFM) states, while they allow for two possibilities in spatial arrangements in simple lattice geometries, there is greater freedom for the symmetry breaking in the pseudospin space. In particular, as shown in Refs.~\cite{Ina2013MPL,Sotnikov2014PRA}, for the SU(3)-symmetric mixture we observe two types of antiferromagnetic ordering: a color-density wave (CDW2) and color-selective antiferromagnetic (CSAF) state. In particular, the spontaneous breaking of the symmetry in the direction of the third spin component (again, restricting in measurements to easy axes) means that the CDW2 and CSAF states are characterized by $m^{(3)}_{A,B}=0$ with $m^{(8)}_{A,B}\neq0$ and $m^{(3)}_A=-m^{(3)}_B\neq0$ with $m^{(8)}_A=m^{(8)}_B\geq0$, respectively; see also Eq.~(\ref{eq.magn}) and Fig.~\ref{fig4}.

\begin{figure}
\includegraphics[width=\linewidth]{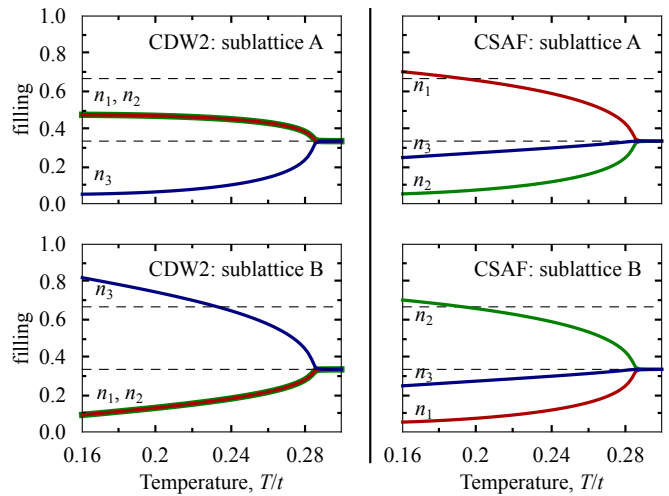}
    \caption{\label{fig4} (Color online)
    Temperature dependence of the relative occupation of two sublattices by spin components for two possible states in the 2s-AFM phase: CDW2 (left) and CSAF (right) at the fixed interaction strength $U/t=12$.}
\end{figure}
It should be mentioned that in our analysis of the 2s-AFM states we also observe the presence of the net magnetization along the direction chosen by the symmetry-breaking mechanism. Moreover, from Fig.~\ref{fig4} one can conclude that it has opposite signs in the CDW2 and CSAF states (i.e., $n_{3A}+n_{3B}>2/3$ and $n_{3A}+n_{3B}<2/3$, respectively) and increases with the temperature decrease. This means that at low temperatures the SU(3)-symmetric mixtures with the fixed equal number of atoms in three different hyperfine states can become unstable towards phase separation into the CDW2- and CSAF-ordered domains that compensate for the residual net magnetization produced by each other. Note that, according to Fig.~\ref{fig4}, the boundary between 2s-AFM and paramagnetic (PM) states in Fig.~\ref{fig2} (the line $T_{c2}$) corresponds to the second-order phase transition similarly to two-component mixtures.

Concerning other phase boundaries in Fig.~\ref{fig2}, it is necessary to note that the transition lines are obtained from the corresponding analysis of the magnetizations (\ref{eq.magn}). At the critical temperature~$T_{c1}$ we observe discontinuities in these observables, i.e., the first-order phase transitions from the 3s-AFM state to both PM and 2s-AFM states. It should be mentioned that, regarding the explicit determination of the phase transition line in this case, it is necessary to analyze the grand-canonical potentials and thus study the system in more detail for possible coexistence regions and metastable solutions. Here we must stress that in the case of a nonbipartite sublattice structure, this task becomes highly nontrivial, thus transforming into a separate problem. However, the corresponding analysis of the impurity problem~(\ref{HamAIM}) can be performed within the ED solver with no significant effort since one has direct access to all eigenstates and partition functions that are also used in Eq.~(\ref{eq.GF}). Hence, from these estimates we can conclude that there is no evidence of the coexistence between PM and 3s-AFM phases, but, at the same time, there are signatures that a narrow coexistence region (with a height of $\Delta T\sim0.02t$) between 2s- and 3s-AFM phases can be present. Therefore, the actual transition line for $T_{c1}$ can be expected to be lower from the side of the 2s-AFM phase than it is depicted in Fig.~\ref{fig2} (i.e., the line shown is effectively the upper boundary for the 3s-AFM phase). Let us emphasize that for a correct determination of the transition line one must solve a separate problem and analyze the grand-canonical potentials corresponding to the original lattice Hamiltonian~(\ref{HubbSU3}) in different phases with a proper account of the sublattice structure.

\begin{figure}
\includegraphics[width=\linewidth]{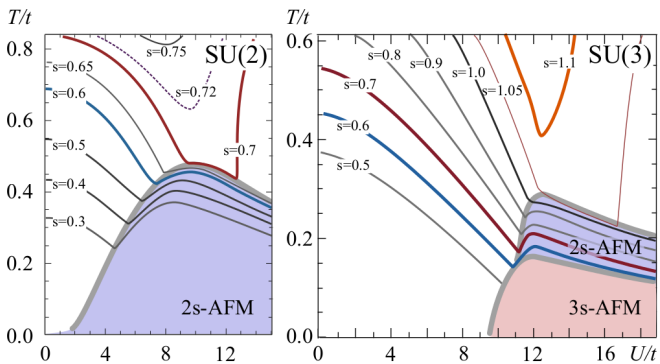}
    \caption{\label{fig5} (Color online)
    Isentropic lines and magnetically ordered phases of the SU(2)- and SU(3)-symmetric Hubbard models obtained by DMFT.}
\end{figure}

Finally, we study the entropy of the system depending on the coupling strength $U/t$ and the temperature $T/t$ at $n=1$. This allows us not only to estimate the critical entropies necessary for approaching the magnetically ordered phases specified in Figs.~\ref{fig2} and \ref{fig3}, but also to study an increase in the magnitude of the Pomeranchuk effect peculiar for the multiflavor mixtures. For this purpose, in Fig.~\ref{fig5}
we also show the isentropic lines and the phase diagram for the SU(2)-symmetric mixture (the corresponding data are taken from Ref.~\cite{Sotnikov2012PRL}) that allow us to make a direct comparison between two- and three-component ultracold atomic mixtures in optical lattices from that perspective.

From Fig.~\ref{fig5} we conclude that, despite the stronger suppression of the antiferromagnetic phases in units of the temperature $T/t$, the situation looks very optimistic for three-component mixtures from the point of view of the critical entropies necessary for approaching quantum magnetism based on the superexchange mechanism. In particular, there is a significant increase (approximately by 50\%) of the critical entropy necessary for approaching 2s-AFM states that we attribute to the stronger Pomeranchuk effect. Of course, one should remember that DMFT provides inaccurate results for the values of the critical entropy in two-component mixtures due to its limitation on local correlations (see Ref.~\cite{Fuchs2011PRL} for more accurate data from exact methods). Naturally, the inaccuracy of the same origin should be expected in our estimates for the SU(3)-symmetric mixture. However, the magnitude of the effect allows us to conclude that the advantageous properties should be confirmed with other theoretical methods and observed in direct experiments. According to our analysis, it is also a good sign that the critical entropy for the 3s-AFM phase $s_{c1}^{\text{DMFT}}\lesssim0.5$ is of a magnitude that is realistic to approach according to the current and expected progress in cooling techniques for ultracold fermionic mixtures.

\section{Conclusion and Outlook}

We developed a theoretical approach that allows us to study in detail the magnetic phase diagram of three-component fermionic mixtures in optical lattices with simple cubic geometry at finite temperature and a filling of one particle per site. At weak coupling for the SU(3)-symmetric mixture we observed a complete suppression of magnetic phases that arise then only in the region of the corresponding crossover to the Mott-insulating state. Within the two-sublattice and real-space DMFT generalizations used we studied two- and three-sublattice antiferromagnetically ordered states and analyzed their structure and critical temperatures. 
Despite the more complex structure of magnetic phases, in the entropy analysis we identified possible advantages of these mixtures in comparison with the two-component counterparts for approaching quantum magnetism in optical lattices due to more pronounced cooling that is facilitated by the stronger Pomeranchuk effect.

From a theoretical point of view, many interesting questions are important to study further in this field of research. In particular, due to limitations of our approach, we were unable to make a strong statement regarding the quantum critical point at $T=0$, i.e., whether it takes place at $U_c\approx9.6t$ (in accordance with the slope of the critical line in Figs.~\ref{fig2} and \ref{fig5}) or there is an exponential suppression of the critical temperature similar to two-component mixtures. It seems that for this purpose DMFT with the corresponding generalization of the numerical renormalization-group solver \cite{Bul2008RMP} to three-component mixtures that operates at $T=0$ can be considered as a good option. Another interesting direction for theoretical investigations is a more detailed analysis of coexistence regions and phase-separation effects both between 3s- and 2s-AFM states and inside the 2s-AFM phase (i.e., between CDW2 and CSAF states).

Note also that in our calculations we restricted ourselves to the easy-axis directions in the eight-dimensional pseudospin space. Naturally, these configurations are not always preferred by the system consisting of ultracold atoms in different hyperfine states (see, e.g., Ref.~\cite{Sotnikov2013PRA} for two-component mixtures). In particular, as mentioned in Ref.~\cite{Sotnikov2014PRA}, in the case of breaking of the SU(3) symmetry by population imbalance, additional terms appear (analogous to external magnetic fields along easy axes) in the effective Hamiltonian that push the system towards canted magnetic configurations. Therefore, this effect must be properly accounted for in experiments (e.g., by introducing additional rf-pulse rotations in measurements of the magnetic phases studied). Note also that the easy-axis directions should be more feasible in systems with the imbalance in hopping amplitudes. The latter can be successfully realized not only in mixtures of alkaline-earth atoms ($^{173}$Yb or $^{87}$Sr) consisting of atoms in the ground and long-living metastable excited states, but also in $^{40}$K mixtures, where a significant hopping imbalance can be effectively induced by lattice shaking with low heating rate \cite{Jotzu2015PRL}. Therefore, this would open a new direction in theoretical studies of possible advantages of the mass-imbalanced mixtures for the purpose of observations of magnetic phases studied in this paper.

Another important direction concerns the inhomogeneity and finite-size effects in these systems. It is stimulated not only by existing requirements from the experimental side, but also by theoretically studied effects of exotic Mott states \cite{Ina2013MPL} and spin separation of three-component mixtures in the trap \cite{Sotnikov2014PRA} that could lead to finding new mechanisms in cooling of fermionic mixtures in optical lattices. 
Note that for the magnetically ordered states under study we assumed that the condition of one particle per lattice site must be fulfilled. However, analogously to two-component mixtures, this condition does not need to be fulfilled exactly, since antiferromagnetic states are also observed at nonzero doping (e.g., the 3s-AFM state is stable for $n\in[0.98,1.02]$ at $U=14t$ and $T=0.12t$). Also, due to the fact that all the AFM states studied are accompanied by the Mott gap, there is a significant stability region for chemical potentials (e.g., the 3s-AFM state is stable for $\mu\in[0.3U,0.6U]$ at $U=14t$ and $T=0.12t$). The latter allows us to expect magnetically ordered domains of a sufficient size (with the linear size in one spatial direction of the order of ten sites or more, depending on the trap curvature) in the trapped systems, both theoretically and experimentally.

Finally, the approach developed has the necessary requirements in accuracy for a detailed analysis of four-component mixtures. These systems are interesting, in particular, due to ongoing debates concerning theoretical determination of magnetically ordered states at a filling of one particle per site in the Hubbard SU(4)-symmetric model and the corresponding Heisenberg limit (see Refs.~\cite{FAs2005PRB,Par2007JP,Cor2011PRL,Szi2011EPL,Cai2013PRB}). Moreover, the detailed structure of the phase diagram at half filling is not completely determined \cite{Zhou2014PRB}, thus the developed DMFT approach could be useful there as well. By means of four-component ultracold atomic mixtures in optical lattices, possibilities also arise to study from new perspectives the Kondo-lattice model as well as other multiband models with an analog of Hund's coupling in solid-state materials. The latter can be realized now by tuning the relative amplitude of spin-exchange processes between particular spin components in these mixtures \cite{Kra2012Nat,Kra2014S,Scazza2014NP}.

\begin{acknowledgments}
The author thanks Agnieszka Cichy, Anna Golubeva, Walter Hofstetter and Yurii Slyusarenko for continuous support and fruitful discussions. Funding from the German Science Foundation DFG via Sonderforschungsbereich Grant No. SFB/TR 49 during the initial stages of this work is gratefully acknowledged. 
\end{acknowledgments}

\bibliography{A20}% Produces the bibliography via BibTeX.

\end{document}